\documentstyle[psfig]{elsart}
\begin{document}
\begin{frontmatter}
\title{Particle production from off-shell nucleons}
\author{P. Bo\.zek}
\address{NSCL, Michigan State University, MI-48824, USA \\ and \\
Institute of Nuclear Physics, PL-31-342 Krak\'ow, Poland}

\begin{abstract}
Particle production in  equilibrium and nonequilibrium 
quantum systems is calculated.
 The effects of the off-shell propagation of nucleons in medium
on the particle production are discussed. Comparision to the semiclassical 
production rate is given.
\end{abstract}
\end{frontmatter}

The production of mesons and hard photons in intermediate energy heavy ion 
collisions is considered to be sensitive to the collision dynamics and to 
medium effects. The medium effects can be two-fold. First, the properties
 of the mesons produced and propagating in the high density region created 
in the collision can be modified. Also particle production rates can be 
modified by the fact that the colliding nucleons  
 are interacting  with the surrounding  nuclear medium.
 This can lead to a different 
production rate obtained form nucleon collisions  in dense  medium then in free
nucleon-nucleon collisions.

In the work \cite{m1} we presented the results of a calculation of the 
meson production rate in a nonequilibrium system of interacting nucleons.
The nonequilibrium dynamics is described by the one-body Kadanoff-Baym
 equations
for the real-time nonequilibrium  fermion Green's functions. 
The numerical solution is obtained for a spatially homogeneous system
with the initial momentum distribution corresponding two Fermi spheres,
following \cite{pawel}.
The quantum rate of the production of mesond is then calculated using the 
corresponding Kadanoff-Baym equations for  mesons with the one-loop 
self-energy. The meson production rate takes the form~:
\begin{eqnarray}
\label{row_mes}
\frac{d N(p,t)}{d^3 p d t} & = & 2 {\cal R}e \Bigg( - \int_{t_0}^{t} dt^{'}
\Pi^{<}(p,t,t^{'}) D^{>}_0(p,t^{'},t) \nonumber \\ & &
+ \int_{t_0}^{t_0-i\tau_0} dt^{'}
\Pi^{<}(p,t,t^{'}) D^{>}_0(p,t^{'},t) \Bigg) \ ,
\end{eqnarray}

\begin{figure}%\epsfxsize=8cm
%\epsffile{z01.eps}
\parbox[t]{6cm}{
\psfig{figure=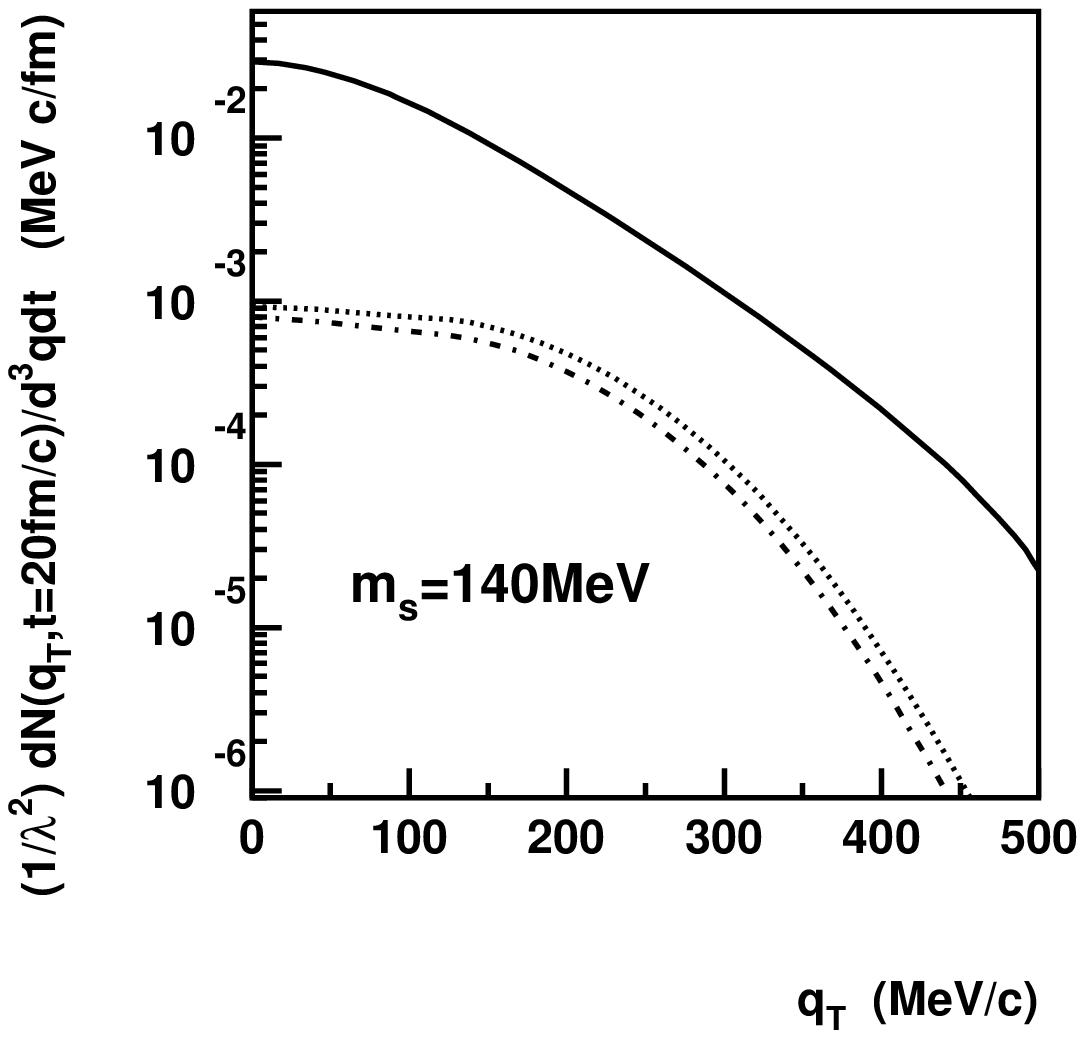,width=6cm,height=6cm,angle=0}}
\hspace{1ex}
\parbox[t]{6cm}{
\psfig{figure=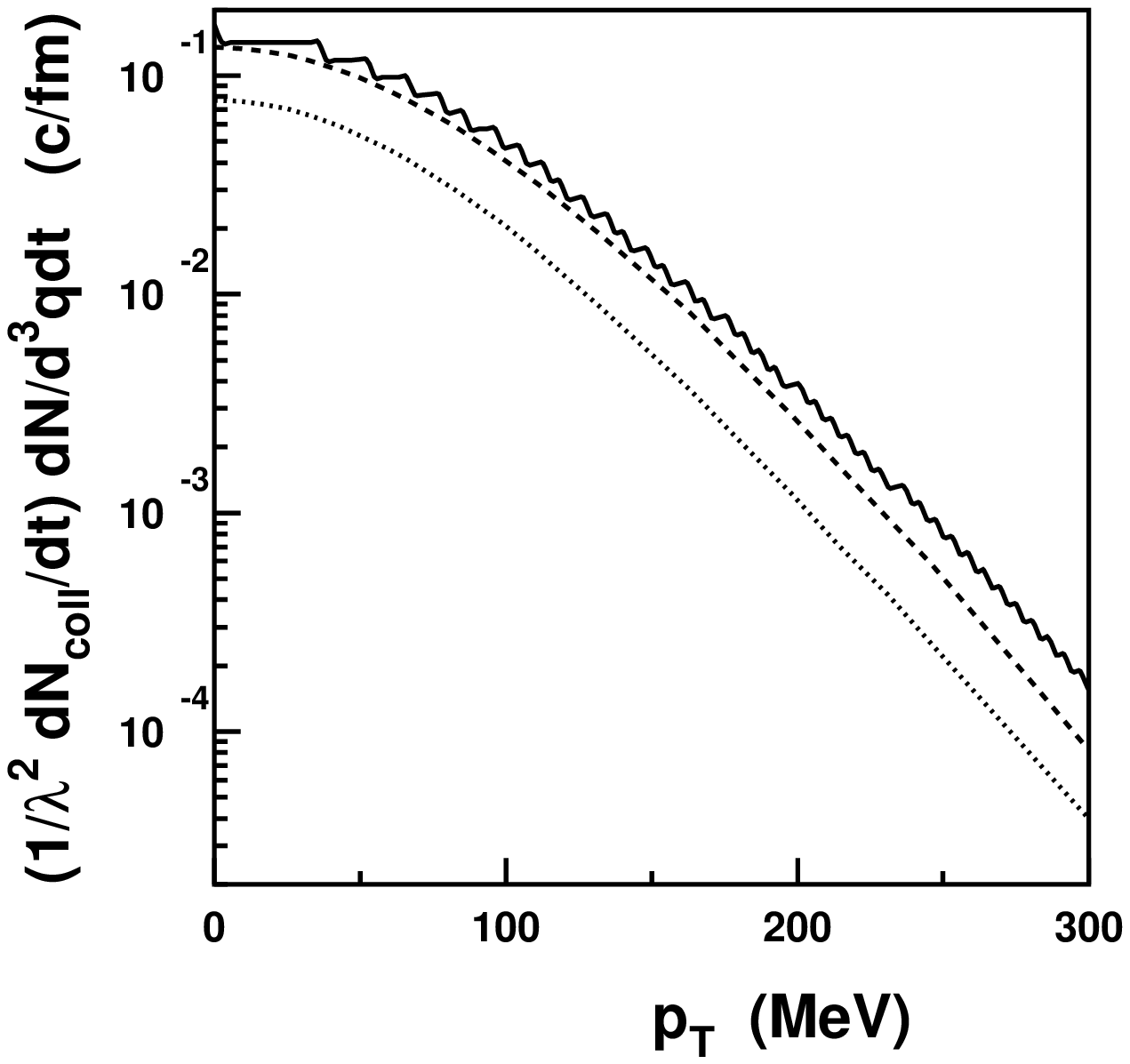,width=6cm,height=6cm,angle=0}}
%\vspace{-4ex}
\caption{\label{1} (left panel) The meson production rate at $t=20$~fm/$c$
from the nonequilibrium evolution at the collision energy of $180$~MeV.
 The solid line denotes the quantum
 production rate. The dotted line is the semiclassical production rate using 
the fermion momentum distribution from the quantum evolution and the 
dashed-dotted line is the one using the momentum distributions obtained from a
 BUU calculation.
(right panel) Production rate of mesons of mass 
$140$~MeV at equilibrium.
 Quantum, semiclassical and semiclassical with unperturbed momentum 
distribution are presented with solid, dashed and dotted lines respectively. }
\end{figure}
where the meson one-loop self energy is given by the fermion Green's functions
 $G$~:
\begin{equation}
\label{se_ol}
\Pi^{<}(p,t_1,t_2)=  - i \lambda^2 \int \frac{d^3q}{(2\pi)^3}
G^{<}(p-q,t_1,t_2)G^{>}(q,t_2,t_1) \ .
\end{equation}
The interaction strength $\lambda$ can in general include momentum 
dependence trough form factors and derivative couplings. In the following
we will compare
the quantum and semiclassical  production rates divided by $\lambda^2$.
The semiclassical production rate was obtained by cutting the meson 
one-loop self-energy diagram with self-energy insertion on the fermion line. 
The matrix element for the production of a meson with momentum $q$ and
 energy $\Omega_q$
 can be written using the retarded fermion Green's functions $G^R$~:
\begin{eqnarray}
\label{mat}
|M|^2=4 \lambda^2 V(p_2-p_4)^2 & \bigg( &|G^R(\omega_{p_1}-\Omega_q,p_1-q)|^2
\nonumber \\ & +&
|G^R(\omega_{p_3}+\Omega_q,p_3+q)|^2 \bigg) \ ,
\end{eqnarray}
where two nucleons with initial momenta $p_1$, $p_2$ scatter into momenta $p_3$
 and $p_4$.
In Ref. \cite{m1} we used a very simple form of the retarded propagator
 including only a constant imaginary fermion self-energy.

In order to study the different approaches to the production rates more 
carefully we repeated the calculation in thermal equilibrium.
The numerical effort is much reduced, allowing for  more precise results.
Also we notice \cite{m1} that it is very difficult to define a nonequilibrium 
quantum production rate because a time average is needed in order to define 
this quantity.
The equilibrium finite temperature system is solved by iterating the
 real-time relations between the fermion Green's functions and
the spectral density, where the spectral density was given by the
Born self-energy depending itself on the fermion Green's functions ($T=20$MeV, 
$\mu=30$MeV, the chemical potential can also include a momentum independent
 mean-field).
The quantum meson production rate is calculated from the one-loop meson
self energy~:
\begin{equation}
\label{eq_mes}
\frac{d N(q)}{d^3 q d t}  =   {\cal I}m  \Pi^{<}(\Omega_q,q)
\end{equation}
 and is scaled by the number of collisions~:
\begin{equation}
\frac{d N_{coll}}{d t} = \int\frac{d^3 p}{(2 \pi)^3}\frac{d \omega}
{2 \pi} G^{<}(\omega,p) \Sigma^{>}(\omega,p) \ ,
\end{equation}
where $\Sigma$ is the fermion self-energy.
The semiclassical production rate is calculated using the matrix element
(\ref{mat}) with the retarded Green's function extracted from the quantum 
calculations. The momentum distribution of fermions is taken using the
quasiparticle fermion energies $f(\omega_p)$ or using the free dispersion 
relation $f(p^2/2 m)$ (unperturbed momentum distributions in equilibrium).
Here also the production rate is scaled by the number of nucleon collisions 
in the quasi-particle approximation.
The results show a smaller difference between the semiclassical and quantum 
production rates than the less accurate nonequilibrium calculations. 
One should also note that the system studied in Ref. \cite{m1} is still
 not at equilibrium at $t=20$~fm/$c$. The differences start to be important 
for large momenta of the mesons or for mesons with high mass. Then the 
quantum treatment of the production rates is necessary and can have 
implications for the predictions on subthreshold meson production.

Another important effect of the medium on production rates can be seen for the
 production of soft particles \cite{kv}. The semiclassical production rate in
 vacuum diverges for soft mesons. In medium it leads to very high production
 rates. These are not realistic due to the requirement of a sufficient
 formation time for the produced particles between collisions.
The use of a damping width
 in the retarded propagators regularizes this behavior.
The result is identical to the quantum production rate with quasi-particle 
approximation for cut fermion lines. In the right panel of Fig.2 we show
 the comparison of the quantum production rate and the semiclassical 
production rates using the in medium and the vacuum matrix element. 
The results show the importance of taking correctly the medium effects into
 account when calculating the production rates for soft particles.
We found  that  both the real and the imaginary 
part of the self-energy in the retarded 
fermion propagator inflence  substantially  the results for
the production rate.
\begin{figure}%\epsfxsize=8cm
%\epsffile{z01.eps}
\parbox[t]{6cm}{
\psfig{figure=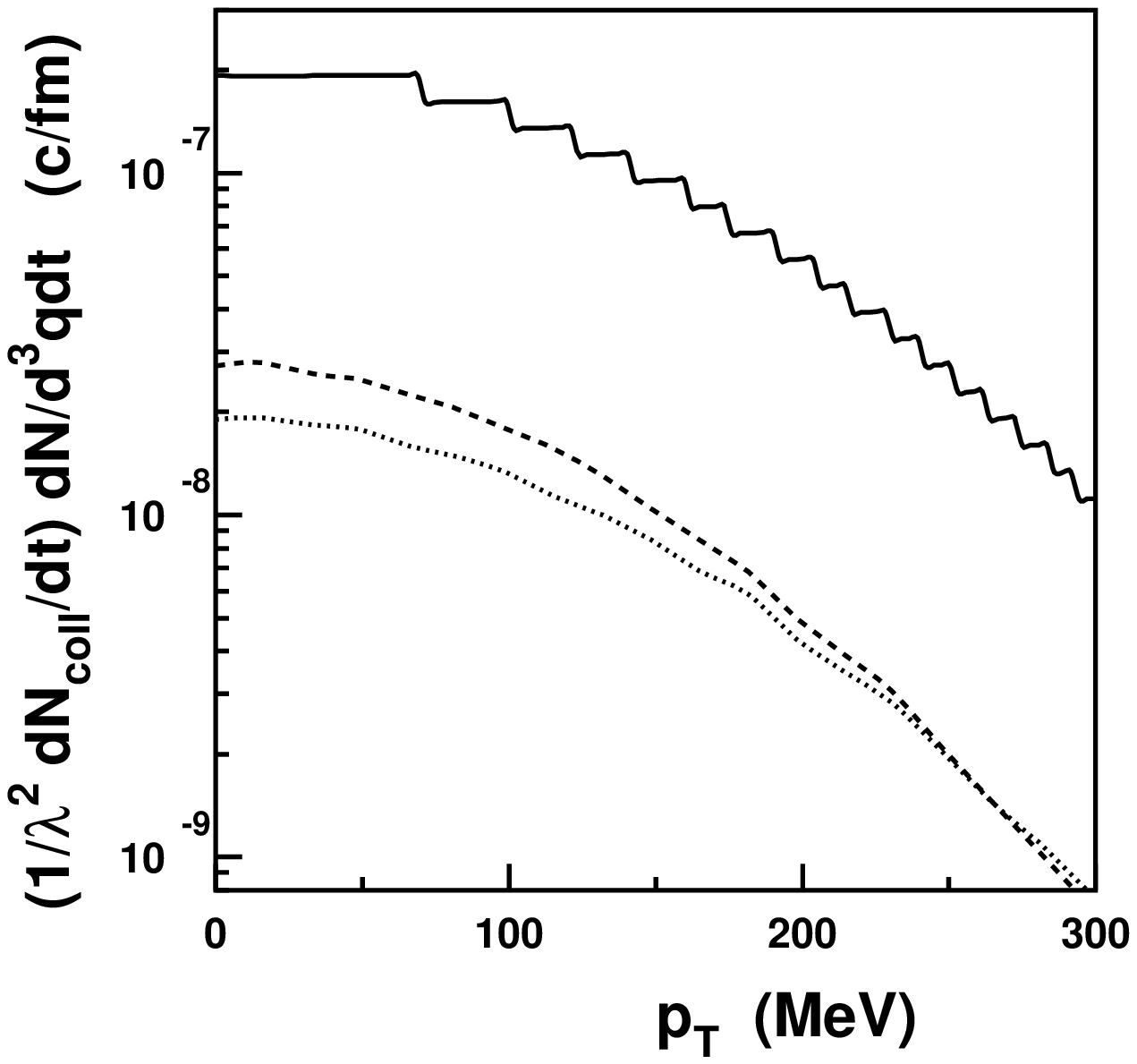,width=6cm,height=6cm,angle=0}}
\hspace{1ex}
\parbox[t]{6cm}{
\psfig{figure=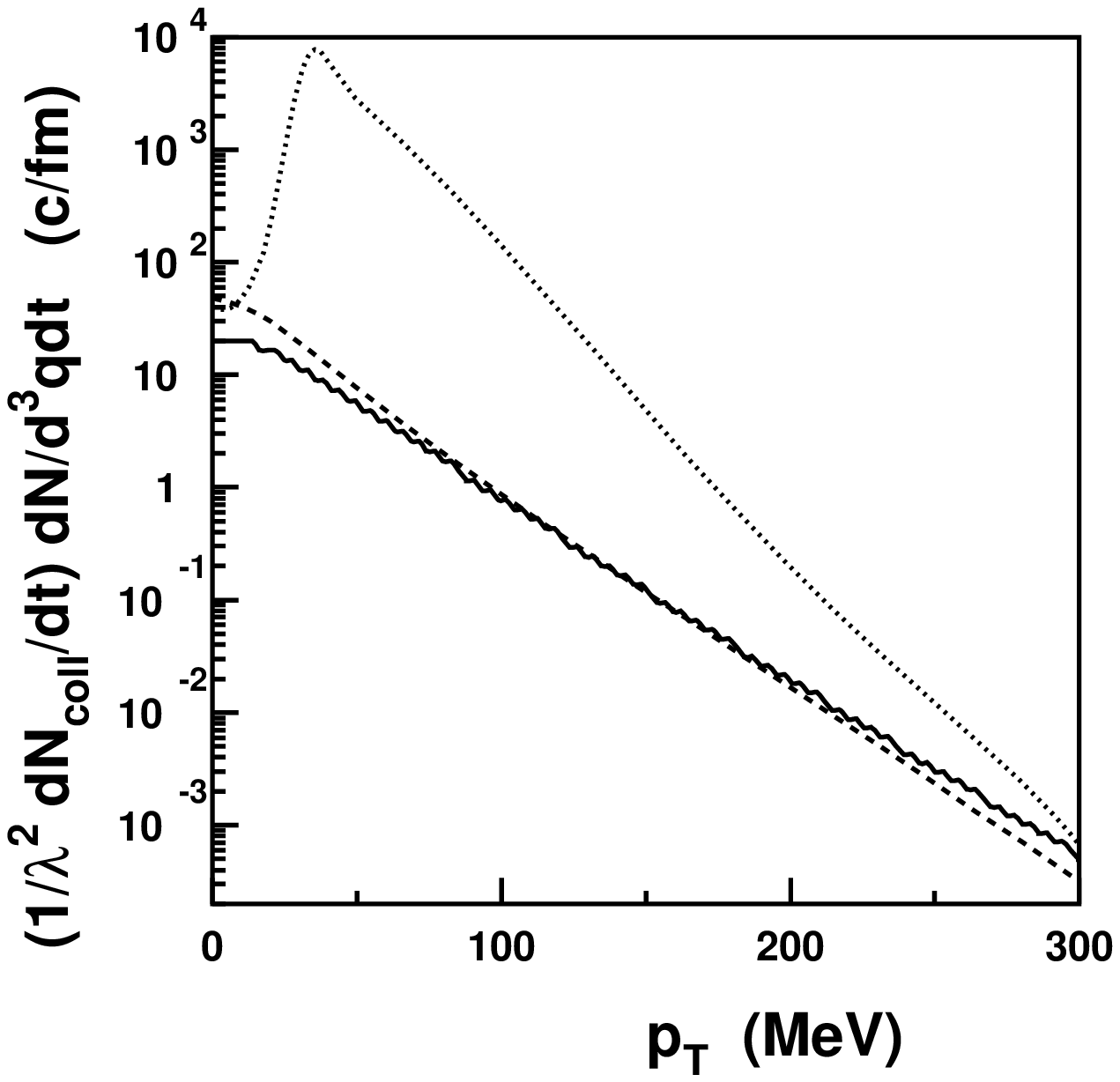,width=6cm,height=6cm,angle=0}}
%\vspace{-4ex}
\caption{\label{2} (left panel) Production rate of mesons of mass 
$500$~MeV at equilibrium.
 Quantum, semiclassical and semiclassical with unperturbed momentum 
distribution are presented with solid, dashed and dotted lines respectively.
(right panel) Production rate of mesons of mass 
$20$~MeV at equilibrium.
 Quantum, semiclassical and semiclassical with vacuum scattering matrix 
distribution are presented with solid, dashed and dotted lines respectively. }
\end{figure}

The calculations using the spectral function in equilibrium are numerical 
fast and accurate. They represent an alternative to the full solution of the 
Kadanoff-Baym equations, which is very difficult  in homogeneous systems and 
impractical for a realistic description of nuclear collision dynamics. Instead
it seems possible to use equations for the local spectral functions,
 equivalent to the gradient expansion of the Kadanoff-Baym equations.

The author would like thank P. Danielewicz and J. Knoll for discussions.
He would like  to express  his gratitude for the hospitality at the
 University of Heidelberg where this work started and at the Michigan
 State University. He also acknowledges the financial support of the Alexander
von Humboldt foundation.

\end{document}